\documentstyle[aas2pp4,psfig]{article}

\def\la{\mathrel{\mathchoice {\vcenter{\offinterlineskip\halign{\hfil
$\displaystyle##$\hfil\cr<\cr\noalign{\vskip1.5pt}\sim\cr}}}
{\vcenter{\offinterlineskip\halign{\hfil$\textstyle##$\hfil\cr<\cr
\noalign{\vskip1.0pt}\sim\cr}}}
{\vcenter{\offinterlineskip\halign{\hfil$\scriptstyle##$\hfil\cr<\cr
\noalign{\vskip0.5pt}\sim\cr}}}
{\vcenter{\offinterlineskip\halign{\hfil$\scriptscriptstyle##$\hfil
\cr<\cr\noalign{\vskip0.5pt}\sim\cr}}}}}
\begin{document}

\title{Discovery of a 57-69 Hz quasi-periodic oscillation in GX13+1}
\author{Jeroen Homan\altaffilmark{1},
        Michiel van der Klis\altaffilmark{1},
        Rudy Wijnands\altaffilmark{1},
        Brian Vaughan\altaffilmark{2},
        Erik Kuulkers\altaffilmark{3}}

\altaffiltext{1}{Astronomical Institute 'Anton Pannekoek', University of Amsterdam, and Center for High Energy Astrophysics,  Kruislaan 403, 1098 SJ, Amsterdam, The Netherlands; homan@astro.uva.nl, michiel@astro.uva.nl, rudy@astro.uva.nl}
\altaffiltext{2}{Space Radiation Laboratory, California Institute of Technology, 220-47 Downs, Pasadena, CA 91125; brian@thor.srl.caltech.edu}
\altaffiltext{3}{Astrophysics, University of Oxford, Nuclear and Astrophysics Laboratory, Keble Road, Oxford OX1 3RH, United Kingdom; e.kuulkers1@physics.oxford.ac.edu}

\begin{abstract}
We report the discovery of a quasi-periodic oscillation (QPO) at
61.0$\pm$1.7 Hz with the Rossi X-Ray Timing Explorer in the low-mass
X-ray binary and persistently bright atoll source GX 13+1 (4U
1811--17).  The QPO had an rms amplitude of 1.7$\pm$0.2\% (2--13.0
keV) and a FWHM of 15.9$\pm$4.2 Hz. Its frequency increased with count
rate and its amplitude increased with photon energy.  In addition a
peaked noise component was found with a cut-off frequency around 2 Hz,
a power law index of around --4, and an rms amplitude of $\sim$1.8\%,
probably the well known atoll source high frequency noise. It was only
found when the QPO was detected. Very low frequency noise was present
with a power law index of $\sim$1, and an rms amplitude of $\sim$4\%. A second observation showed similar variability components.  In
the X-ray color-color diagram the source did not trace out the usual
banana branch, but showed a two branched structure.

This is the first detection of a QPO in one of the four persistently
bright atoll sources in the galactic bulge. We argue that the QPO
properties indicate that it is the same phenomenon as the horizontal
branch oscillations (HBO) in Z sources. That HBO might turn up in the
persistently bright atoll sources was previously suggested on the
basis of the magnetospheric beat frequency model for HBO. We discuss
the properties of the new phenomenon within the framework of this
model.

\end{abstract}


\section{Introduction}
Based on their correlated X-ray timing and spectral behavior the
brightest low-mass X-ray binaries (LMXBs) can be divided into two
groups; the atoll sources and Z sources (Hasinger \& van der Klis
\markcite{hk89}1989, hereafter HK89; van der Klis
\markcite{kl95}1995). Atoll sources show two states: the island and
the banana state, after the tracks they produce in an X-ray
color-color diagram (CD). Z sources on the other hand trace out a Z-like shape
in a CD, with usually three branches: the horizontal, the
normal, and the flaring branch.

The power spectra of atoll sources can be described by two noise
components plus, sometimes, a Lorentzian component to describe
quasi-periodic oscillations (QPOs). The first noise component, the
very low frequency noise (VLFN) has a power law shape
$P\propto\nu^{-\alpha}$, with $1\leq\alpha\leq1.5$. The other, the
high frequency noise (HFN), can be described by a power law with an
exponential cut-off $P\propto\nu^{-\alpha} e^{-\nu/\nu_{cut}}$,
usually with $0\leq\alpha\leq0.8$ and $0.3\leq\nu_{cut}\leq25$ Hz.
The HFN sometimes has a local maximum (``peaked noise'') around 10-20
Hz (see van der Klis 1995). Yoshida et al. (1993) found peaked noise
around 2 Hz. Several broad QPO(-like) peaks were found with the Rossi
X-ray Timing Explorer (RXTE) around 20 Hz (Strohmayer et
al. \markcite{st96}1996; Ford et al. \markcite{fo97}1997; Yu et
al. \markcite{yu97}1997; Wijnands \& van der Klis
\markcite{wikl97}1997), and one at 67 Hz (simultaneously with one at
20 Hz, see Wijnands et al. \markcite{wi97b}1997b;). 
In addition to QPOs below 100 Hz, QPOs between 300 and 1200 Hz, the
so-called kHz QPOs, have been found.

The power spectra of Z sources show three broad noise components:
VLFN with $1.5\leq\alpha\leq2$, HFN with $\alpha\sim0$ and
$30\leq\nu_{cut}\leq100$ Hz, and low frequency noise (LFN), which has
the same functional shape as the HFN with $\alpha\sim0$ and
$2\leq\nu_{cut}\leq20$ Hz. Note that despite having the same name,
HFN in Z sources is not the same phenomenon as HFN in atoll sources. Z
sources show three types of QPOs: the normal/flaring branch QPO
(N/FBO) with centroid frequencies from 6 to 20 Hz, the horizontal
branch QPO (HBO) from 15 to 60 Hz, and the kHz QPOs in the same range
as observed in atoll sources. (For an extensive review on the power
spectra of atoll and Z sources we refer to van der Klis
\markcite{kl95}1995. For kHz QPOs  we refer to van der Klis
\markcite{kl97}1997.)

GX 13+1 has been classified as an atoll source, although of all atoll
sources it shows properties which are closest to that seen in the Z
sources (HK89). Moreover, Schulz, Hasinger \& Tr\"umper\markcite{sc89}
(1989) put GX 13+1 among the luminous sources that have been
classified as Z sources. Together with GX 3+1, GX 9+1, and GX 9+9, GX
13+1 forms the subclass of the persistently bright atoll sources. In
the CD they have only been seen to trace out banana branches and their
power spectra can be described by relatively strong ($\sim$3.5\% rms)
VLFN and weak ($\sim$2.5\% rms) HFN, as compared to other atoll
sources. No QPOs have been found before in these sources, neither at
frequencies below 100 Hz nor at kHz frequencies (Wijnands, van der
Klis \& van Paradijs\markcite{wi97a} 1997a; Strohmayer et
al.\markcite{st97} 1997).
For GX 13+1, so far no HFN has been observed. Its banana branch
resembled a more or less straight strip in the CD, whereas the other
three sources showed more curved banana branches (HK89). Stella, White \&
Taylor \markcite{st85}(1985) have reported bimodal behavior of GX 13+1
in the hardness-intensity diagram (HID). In one state the spectral
hardness was correlated with count rate, while in the other it was
anticorrelated. The transition between the two states occurred within
one hour.

The main difference between atoll and Z sources is the mass accretion
rate, $\dot{M}$. Atoll sources accrete at
$\dot{M}\la0.5\dot{M}_{Edd}$, whereas Z sources accrete at near
Eddington rates.  It was proposed by HK89 that a second difference
lies in the strength of the magnetic field strength $B$ of the
accreting neutron star. Recent spectral modeling by Psaltis \& Lamb
\markcite{ps96}(1996) suggests that indeed atoll sources have $B<10^9$
Gauss and Z sources  $B\sim10^9$--$10^{10}$ Gauss. The bright atoll
sources are found to have a higher inferred $B$ than the
low luminosity atoll sources, making them the best candidates to show
Z source HBO.

In this Letter we report the  discovery of a 57--69 Hz  QPO and of a two
branched structure in  the CD and HID of  GX 13+1. We suggest that the
QPO is the same phenomenon as Z source HBO.

\section{Observations and analysis}
We observed GX 13+1 with the proportional counter array (PCA) onboard
RXTE, on October 28 1996 02:15--05:15 UTC and on November 10 1996
09:19--12:13 UTC. Except for a $\sim$300 s interval during the October
28 observation, when proportional counter unit (PCU) four was
inactive, data (a total of $\sim$$1.2\times10^4~$s) were collected
with all five PCUs in three simultaneous data modes: 16 s time
resolution in 129 photon energy bands (covering the energy range 2--60 keV);
$2^{-12}$ s time resolution in 3 bands (2--4.9 keV, 4.9--8.6
keV, and 8.6--13.0 keV); $2^{-16}$ s time resolution in 64 bands
(covering the range 13.0--60 keV).

The 16 s data were used to construct lightcurves (Fig. 1) and CDs and
HIDs (Fig. 2). Only data with all five PCUs on were used. The data
were background corrected, but no dead-time corrections ($\sim$1\%)
were applied. The data gaps in Figure 1 are due to Earth occultations
of the source, or to passages of the satellite through the South
Atlantic Anomaly. The average 2--19.7 keV count rate during the first
part of the October 28 observation is $\sim$3800 cts/s, during the
second and third part $\sim$4700 cts/s, and during the November 10
observation $\sim$4400 cts/s. For the soft color we used the count
rate ratio between 2--3.9 keV and 3.9--6.4 keV; for the hard color the
ratio between 8.6--19.7 keV and 6.4--8.6 keV. For the intensity we
used the count rate in the 2--19.7 keV energy range.

Power density spectra were made of all the $2^{-12}$ s data using 16 s
data segments. For measuring the (low frequency) QPO we fitted the
0.1--256 Hz power spectra with a constant representing the Poisson
level, a power law representing the VLFN, a power law with an
exponential cut-off representing the HFN, and a Lorentzian representing
the QPO. Errors on the fit parameters were determined using
$\Delta\chi^2=1$, upper limits with $\Delta\chi^2=2.71$, corresponding
to a 95\% confidence level. Upper limits on (sub-)harmonics of the QPO
were determined by setting the frequency to half or twice the QPO
frequency and by fixing the FWHM to half or twice the FWHM of the
QPO. Upper limits for kHz QPOs were determined by fitting the
200--2048 Hz power spectra with a constant and a Lorentzian with a
conservatively assumed FWHM of 150 Hz. This was done in the 2--13.0
keV, 13.0--60 keV, and 2--60 keV energy ranges.

\section{Results}
During the October 28 observation the source did not trace out a
banana branch in the CD. Instead two distinct branches could be
identified (see Fig. 2a): a lower branch and an upper branch,
separated by a gap in the CD around a hard color of $\sim$0.6,
corresponding to the first gap in the light curve (Fig. 1). The other
gaps in the light curve did not appear as gaps in the CD and HID.  The
source started at the right end of the lower branch, then moved up to
the top of the upper branch along the curve and ended halfway down the
upper branch. Just before the first gap in the lightcurve the source
was in the upward curved part of the lower branch.  Hence the points
at the left part of the lower branch above hard colors of $\sim$0.5
are in all likelihood evidence of motion towards the upper branch that
continued during the gap. A slightly curved branch was traced out in
the CD during the November 10 observation, close to the upper branch
of the October 28 observation. The source started in the lower part of
the branch, moved to the top and ended in the lower part.
 
In all the October 28 data combined we discovered a QPO at
61.0$\pm$1.7 Hz, with a FWHM of 15.9$\pm$4.2 Hz, an rms amplitude of
1.7$\pm$0.2\% and a significance of 4.8$\sigma$ (2--13.0 keV). The
power spectrum is shown in Figure 3. In the lower branch the QPO could
not be detected significantly (2.2$\sigma$) at 61.0 Hz, with an upper
limit on the rms amplitude of 1.7\%. In the upper branch the QPO was
detected (4.7$\sigma$) at 59.7$\pm$1.9 Hz with an rms amplitude of
2.1$^{+0.3}_{-0.2}$\% and a FWHM of 21.4$\pm$5.8 Hz. A count rate
selection showed that the QPO frequency on the upper branch increased
with count rate; from 57.0$\pm$1.4 Hz between $\sim$4050 and
$\sim$4700 cts/s to 64.5$\pm$2.1 Hz between $\sim$4700 and $\sim$5600
cts/s. The FWHM and rms amplitude did not change significantly; from
11.6$\pm$4.4 Hz to 14.1$\pm$7.8 Hz and from 1.8$^{+0.3}_{-0.2}$\% to
2.0$^{+0.4}_{-0.3}$\% respectively. Upper limits on a subharmonic or
second harmonic were respectively 1.0\% rms and 0.8\% rms. The VLFN
had an rms amplitude of $\sim$4.3\% and a slope of $\sim$1.3. Its
properties did not vary significantly between the two branches. In
addition to the VLFN a peaked noise component was detected
(7.4$\sigma$) with a cut-off frequency of $2.1\pm0.6$ Hz, a power law
index of --3.6$\pm$1.2, and an rms amplitude of
1.9$^{+0.2}_{-0.1}$\%. This noise feature is probably atoll source
HFN. In the upper branch it had an rms amplitude of 2.3$\pm$0.2\%, in
the lower branch it was undetectable with an upper limit of 1.5\%
rms. No kHz QPOs were found between 200 and 2048 Hz, with upper limits
on the rms of 2.1\% (2--13.0 keV), 21.0\% (13.0--60 keV), and 2.3\%
(2--60 keV).

The strength of the QPO increased with photon energy (Fig. 4). The
energy spectrum of the HFN was both consistent with that seen for the
QPO and with being constant. The VLFN strength increased up to
$\sim$11 keV and decreased thereafter.

In the November 10 data we may have detected (2.5$\sigma$) a similar
QPO at 71.0$\pm$2.5 Hz with a FWHM of 12.4$^{+11.3}_{-5.9}$ Hz and an
rms amplitude of 1.2$^{+0.8}_{-0.4}$\%. At high count rates,
$\sim$4500 cts/s, the QPO was found to be a bit more significant
(3.1$\sigma$) at 68.8$\pm$2.0 Hz, with an rms amplitude of
1.4$^{+0.3}_{-0.2}$\% and a FWHM of 10.4$\pm$4.9 Hz. At lower count
rates, $\sim$4100 cts/s, an upper limit for the rms amplitude was
obtained of 1.2\%. The VLFN had the same slope as in in the first
observation, and an rms amplitude of $\sim$3.7\%. Again HFN was
detected (6.4$\sigma$) with an rms amplitude of 1.7$\pm $0.2\%, a
power law index of -4.5$\pm$1.5, and a cut-off frequency of
$2.1\pm0.9$ Hz. Again no kHz QPOs were detected, with upper limits on
the rms of 2.3\% (2--13.0 keV), 19.9\% (13.0--60 keV), and 2.2\%
(2--60 keV).

\section{Discussion}
We have discovered a QPO in GX 13+1 between 57 and 65 Hz and around 69
Hz. It is the first time a QPO has been found in a persistently bright
atoll source. The rms amplitude of the QPO increased with photon
energy. Together with the QPO, a peaked noise component was found,
probably atoll source HFN, with a cut-off frequency of $\sim$2 Hz. The
HFN was only detected when the QPO was present. In the October data
the QPO was only found in the upper branch, i.e. at high count
rates. On the upper branch the QPO frequency increased with count rate
from $\sim$57 Hz to $\sim$65 Hz. In the November data the QPO could
only be detected at high count rates. Assuming that $\dot{M}$ and
count rate were positively correlated, the QPO frequency increased
with $\dot{M}$, at least on the upper branch in the October
observation. However, during the November observation we found the QPO
at $\sim$69 Hz; the mean count rate was then lower than during the
October observation of the $\sim$65 Hz QPO.

The pattern traced out by GX 13+1 in the CD and HID in our
observations does not resemble the patterns traced out by other atoll
sources. Atoll sources trace out islands and/or a banana
branch. During EXOSAT observations GX 13+1 traced out a banana branch
(HK89) which is quite different from what we observed with RXTE. We
also do not find evidence for the bimodal behavior found by Stella et
al. \markcite{st85}(1985) in the sense that both our branches in the
HID show a positive correlation between hard color and count rate. The
relatively sharp turn in the CD and HID may be related to one of the
vertices in the patterns traced out by Z sources.

Comparison with EXOSAT observations, using our
RXTE energy spectra folded with the EXOSAT response matrix, shows that
the source was $\sim$30\% brighter during the RXTE
observations. (Note that in the RXTE ASM lightcurve the source shows
intrinsic variations of $\sim$50\%; during  our observations the
ASM count rate was near average.) This might explain why the two
branched structure has not been seen before in GX 13+1. In any case EXOSAT was not sensitive enough to have detected the QPO
reported in this Letter.

Recently Stella \& Vietri \markcite{st98}(1998) proposed that at least
some of the $<100$ Hz QPOs observed in atoll sources are due to
Lense-Thirring precession of the inner part of the accretion disk. In
order to test this model one needs the frequencies of the
simultaneously observed kHz QPOs. Since no kHz QPOs have been observed
in GX 13+1 it is not possible to test this model. The
upper limits on kHz QPOs are comparable with those found in the other
members of the subclass (Wijnands et al. \markcite{wi97a}1997a;
Strohmayer et al. \markcite{st97}1997)

The QPO properties (frequency,  dependence on $\dot{M}$, 
energy spectrum, and the simultaneous presence of a band limited noise
component [the HFN]) are all similar to those found for HBO in Z sources. On
the basis of the above described similarities we suggest that the QPO
we found is the same phenomenon as the HBO, and that the HFN component
is related to the QPO in a similar way as Z source LFN to HBO. The
identification of atoll source HFN with Z source LFN was previously
proposed by van der Klis \markcite{kl94}(1994).

The HBO in Z sources can be explained by the magnetospheric beat
frequency model (Alpar \& Shaham\markcite{al85}, 1985; Lamb et
al.\markcite{la85}, 1985) according to which the observed QPO
frequency is the difference between the neutron star spin frequency
and the frequency at which blobs of matter orbit the neutron star at
the magnetospheric radius. In this model the frequency increases with
$\dot{M}$ and decreases with neutron star spin frequency and $B$.  The
HBO is observed at frequencies between 15 and 60 Hz. LFN is found with
a cut-off frequency between 2 and 20 Hz; it usually appears and
disappears together with the HBO. In quantitative models the LFN is
naturally produced as an extra component to the HBO, with a total
power that is comparable to that in the HBO (Shibazaki \& Lamb
\markcite{sh87}1987). Both components get stronger with photon energy.

According to HK89 the properties of atoll and Z sources are determined
by $\dot{M}$ and $B$. Spectral modeling based on this picture shows
that of all atoll sources, the persistently bright subclass,
especially GX 13+1, have $\dot{M}$ and $B$ closest to those of the Z
sources (Psaltis \& Lamb \markcite{ps96}1996). On the basis of the
magnetospheric beat frequency model one might therefore expect to see
HBO-like phenomena in these sources. The detection of a HBO-like
phenomenon in GX 13+1 seems to confirm these expectations.  The fact
that the QPO frequency in GX 13+1 is at the high end of the HBO
frequency range in Z sources can be explained by a slower spinning
neutron star or/and by a $B$ that is lower than in the Z sources, but
still high enough to produce HBO. A higher $\dot{M}$ than Z sources is
unlikely in view of the luminosity.

\acknowledgments{This work supported in part by the Netherlands Foundation for Research in Astronomy (ASTRON) grant 781-76-017. }

\newpage

\begin{figure}[h]
\begin{center}
\psfig{figure=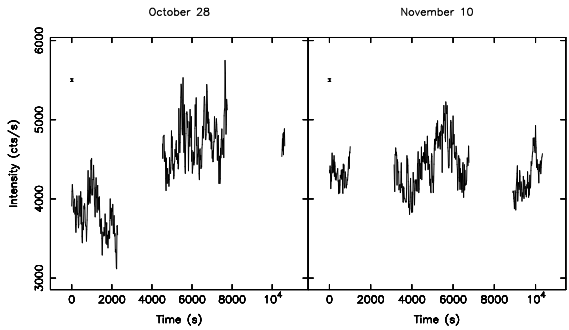,width=8cm}
\caption{The 5-detector light curves of the October 28 and November 10 observations. The intensity is the count rate in the 2--19.7 keV band. Time resolution is 16 s. $T=0$ on October 28 corresponds to 02:16:49 UTC and on November 10 to 09:19:29 UTC. Typical error bars are shown in the upper left of the figures.
\label{lightcurve}}
\end{center}
\end{figure}

\newpage

\begin{figure}[h]
\begin{center}
\psfig{figure=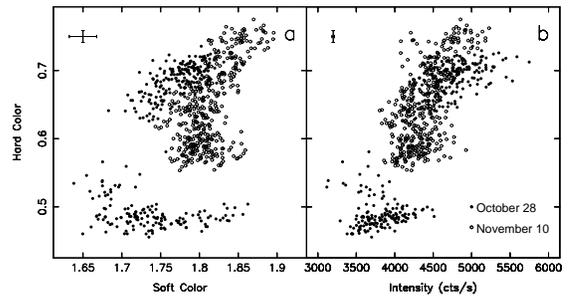,width=8cm}
\caption{The color-color diagram (a) and hardness-intensity diagram (b) of GX 13+1. Filled circles depict  October 28 data, open circles  November 10 data. For energy bands, see text. Each point is the average of a 16 s interval. Typical error bars are shown in the upper left of the figures.
\label{col-hid}}
\end{center}
\end{figure}

\newpage

\begin{figure}[h]
\begin{center}
\psfig{figure=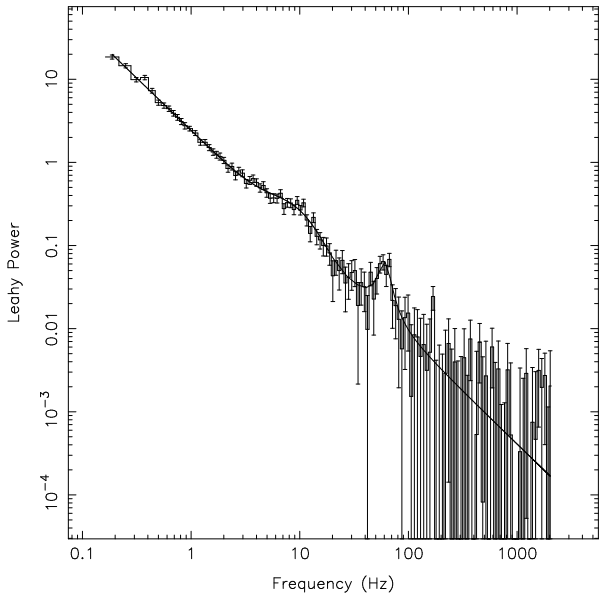,width=8cm}
\caption{The Leahy normalized power spectrum of GX 13+1 on October 28 in the energy range 2--13.0 keV. The Poisson level has been subtracted.\label{power}}
\end{center}
\end{figure}

\newpage

\begin{figure}[h]
\begin{center}
\centerline{\psfig{figure=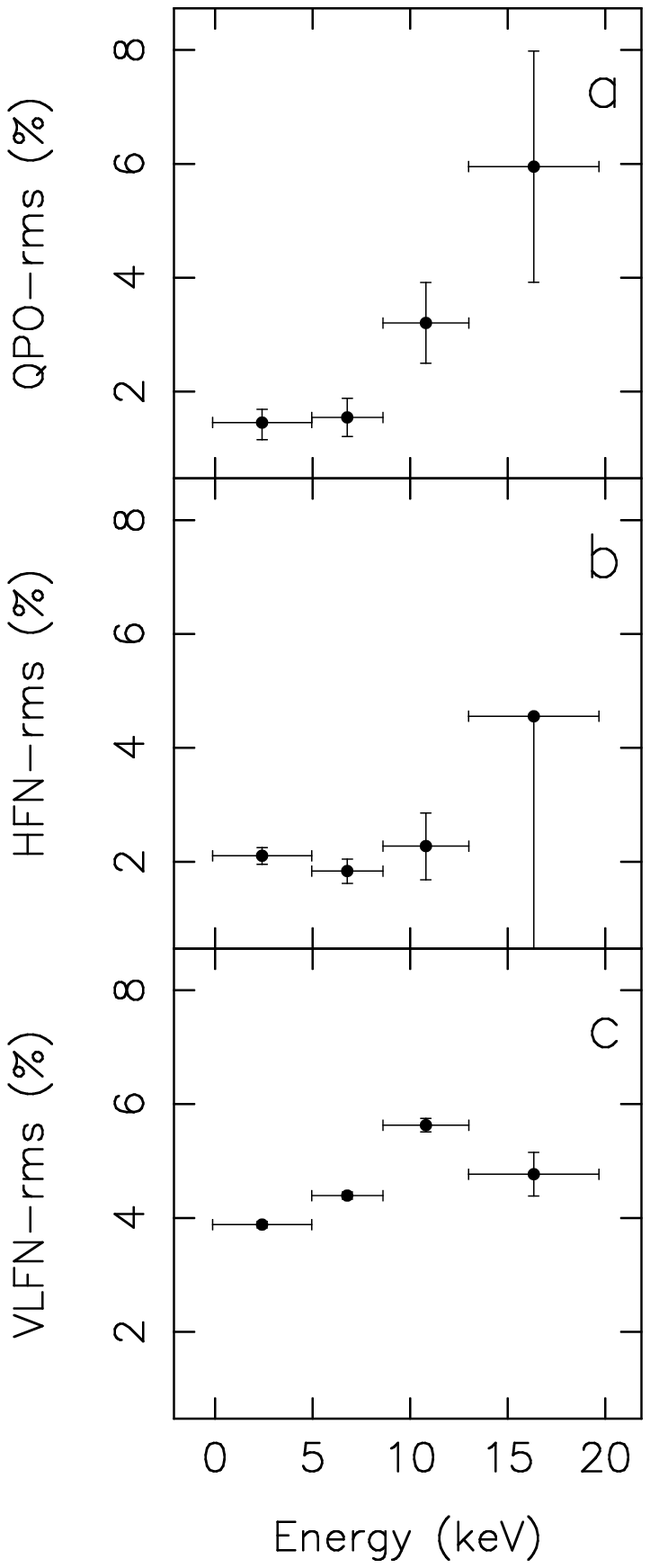,height=8cm}}
\caption{Energy dependence of the QPO (a) the HFN and (b) the VLFN (c). The fourth point in (b) is an upper limit.\label{energy}}
\end{center}
\end{figure}






\begin{references}
\reference{al85} Alpar, M.A., Shaham, J., 1985, \nat, 316, 239
\reference{fo97} Ford, E., Kaaret, P., Tavani, M., Barret, D., Bloser, P., Grindlay, J., Harmon, B.A., Paciesas, W.S., Zhang, S.N., 1997, \apj, 475, 123
\reference{hk89} Hasinger, G., van der Klis, M., 1989, \aap, 225, 79
\reference{la85} Lamb, F.K., Shibazaki, N., Alpar, M.A., Shaham, J., 1985, \nat, 317, 681
\reference{ps96} Psaltis, D., Lamb, F.K., 1996, Poster presented at the NATO ASI: The many faces of neutron stars, Lipari
\reference{sc89} Schulz, N.S., Hasinger, G. and Tr\"{u}mper, J., 1989, \aap, 225, 48
\reference{sh87} Shibazaki, N., Lamb, F.K., 1987, \apj, 318, 767
\reference{st85} Stella, L., White, N.E., Taylor, B.G., 1985, in  Proc. ESA Workshop: {\it Recent Results on Cataclysmic Variables}, Bamberg (ESA SP-236), 125
\reference{st98} Stella, L., Vietri, M., 1998, \apj, 492, 59
\reference{st96} Strohmayer, T.E., Zhang, W., Swank, J.H., Smale, A., Titarchuk, L., Day, C., Lee, U., 1996, \apjl, 469, 9
\reference{st97} Strohmayer T., et al. 1997, poster presented at the 8th Annual October Astrophysics Conference, Maryland ``Accretion processes in Astrophysics: Some like it hot'' 
\reference{kl94} van der Klis, M., 1994, \aap, 283, 469
\reference{kl95} van der Klis, M., 1995, in {\it X-Ray Binaries}, eds. Lewin, W.H.G., van Paradijs, J., van den Heuvel, E.P.J. Cambridge University Press, 252
\reference{kl97} van der Klis, M., 1997, to appear in Proc. NATO Advanced Study Institute "The many faces of neutron stars", Lipari, Italy, 1996, astro-ph/9710016
\reference{wi97a} Wijnands, R., van der Klis, M., van Paradijs, J, 1997a, to appear in the proceedings of IAU Symposium 188 "The Hot Universe", astro-ph/9711222
\reference{wi97b} Wijnands, R., van der Klis, M., M\'{e}ndez, M., van Paradijs, J., Lewin, W.H.G., Lamb, F.K., Vaughan, B., Kuulkers, E., 1997b, accepted \apjl, in press
\reference{wikl97} Wijnands, R.A.D., van der Klis, M., 1997, \apj, 482, 65
\reference{yo93} Yoshida, K., Mitsuda, K., Ebisawa, K., Ueda, Y., Fujimoto, R., Yaqoob, T., Done, C., 1993, \pasj, 45, 605
\reference{yu97} Yu, W., Zhang, S.N., Harmon, B.A., Paciesas, W.S., Robinson, C.R., Grindlay, J.E., Bloser, P., Barret, D., Ford, E.C., Tavani, M., Kaaret, P., 1997, \apj, 490, 153 
\end{references}
\end{document}